\documentstyle[preprint,aps]{revtex}
\begin{document}

\title{The Ba 4d-f giant dipole resonance
as a probe of the structure of\\
endohedral ${\bf {\rm Ba@C}_n}$ metallofullerenes}

\author{Jarek Luberek and G\"oran Wendin}

\address{Department of Applied Physics, Chalmers University of
Technology\\
S-412 96 G\"oteborg, Sweden}
\date{\today}
\maketitle

\begin{abstract}
We calculate the x-ray absorption near edge structure (XANES)
modulating the Ba 4d-photoabsorption cross section - the giant
dipole resonance -  centered around 110 eV photon energy for
several models of spherical and {\em non-spherical} ${\rm Ba@C}_n$
metallofullerene cage systems.
In the cases considered, the XANES interference patterns provide
clear structural "fingerprints", distinguishing between center versus
off-center Ba position and spherical versus deformed  ${\rm C}_n$
shell.
\end{abstract}


\narrowtext

Cage molecules are well known for spectacular interference
effects,
causing very pronounced structure in photoabsorption cross
sections. Nearly ideal cages should be provided by fullerene
molecules
${\rm C}_n,\quad n=60,70,\ldots$, in gas phase or in solid form.
Among various forms of doped
fullerenens, there is the possibility to put metal atoms inside
${\rm C}_n$, endohedral metallofullerenes - ${\rm M@C}_n$
\cite{1}. Recently there
has been a wealth of reports on the
preparation and characterization of endohedral metallofullerenes
${\rm M@C}_n$
\cite{2,3,4}  as well as calculations of structure, affinity and
ionization energies \cite{5,6,7}.

The presence of a heavy metal atom inside a cage of ligands
provides a very interesting many-electron system:
the collective dynamics of the coupled many-electrons systems
may give rise to giant dipole resonances,
and single-particle multiple scattering will lead to
standing wave patterns and resonances inside the cage\cite{WeWa,WaWe}.
As a result, the broad giant dipole resonance
will be modulated by x-ray absorption near edge structure (XANES)
as illustrated by the Ba 4d-photoionization cross section
in solid ${\rm BaF}_2$ and ${\rm YBaCuO}$ (see e.g. ref. \cite{Onell}).

In this Letter we present results from an investigation of the
collective and single-electron response in photoemission from the
4d core level of
a Ba atom placed inside a model fullerene cage in cases of spherical
and {\em non-spherical}
symmetry. We focus attention on two situations: (i) a Ba atom
displaced from
the center of a spherical ${\rm C}_{60}$  model cage,
${\rm Ba@C}_{60}({\rm C}_{{\infty}v})$,
and (ii) a Ba atom at the center of a
non-spherical cage consisting of a short tube with hemispherical
endcaps, to be
referred to as ${\rm Ba@C}_{90}({\rm D}_{{\infty}h})$.
We find that the multiple scattering
structure of the photoabsorption cross section is a sensitive probe
of the position of the
emitter atom and the shape of the cage. We propose that the
oscillations of the 4d-f giant dipole resonance can provide a
chacteristic "fingerprint" of the system -  distinguishing almost
by inspection between center vs off-center position of
the Ba emitter atom as well as spherical vs non-spherical shape of
the cage.

The ${\rm Ba@C}_n$ structure is
defined
by a point charge ${\rm Z}_{\text{Ba}}=56$ for the Ba nucleus
and a
surface effective nuclear charge $Z_{\text{shell}}$ for the $C_n$ shell
\cite{WeWa}.
In this way the nuclear point charges of the
real ${\rm C}_n$ structure is averaged to a homogeneous surface charge
distribution
corresponding to the number of carbon valence electrons
(4 per C-atom). This
is in fact an effective 2-dimensional jellium model. We then fill in
the
appropriate number $N=Z_{\text{Ba}}+Z_{\text{shell}}$
of electrons for a neutral ${\rm Ba@C}_n$ cluster and
calculate the electronic structure selfconsistently within the
local-density approximation (LDA) (the structure and dynamics of
spherical ${\rm La@C}_{60}$ was studied in the same way
by W\"astberg and Wendin \cite{WaWe}).

To find the electronic structure of a {\em non-spherical}
cluster we
expand the initial bound state wave functions
$\psi({\bf r}) = (1/r)\sum_L u_L(r) Y_L(\Omega)$
and the final state Green functions
$G({\bf r},{\bf r}',E) = \sum_{L,L'} Y_L(\Omega)G_{L,L'}(r,r')
Y^*_{L'}(\Omega')$, $L = (l,m)$,
in spherical harmonics around the center of the
heavy central atom \cite{9,11}.
In such a one-center expansion, the molecular
problem reduces to an
atomic problem, but with coupled angular-momentum channels.
The advantage is
that the Coulomb interaction can be expressed
in terms of spherical
harmonics, which allows straigthforward calculation
of the dielectric
response function. Hence electron-electron correlation
and many-particle
dynamics in the
cluster can be studied using well established many-body
techniques for atomic systems.

The resulting coupled radial Schr\"odinger equations were
solved by
procedures described in Ref. \cite{11}.
{}From the spherical
expansion of initial state wave functions
and final state Green
functions \cite{Com1} we
then construct the dielectric susceptibility
$\chi_0({\bf r},{\bf r}';\omega)$
and calculate the induced charge $\delta n({\bf r};\omega)$ in a
self-consistent manner within the TDLDA (time
dependent local density approximation) \cite{ZaSo}.
We finally obtain the
photoabsorption cross section
$\sigma(\omega) \sim \omega {\rm Im}\int z \delta n({\bf r};\omega) dr$
where $z=r \cos \theta $ is the the electric dipole operator
\cite{Com2} with the electric field along the z-axis.
In this work the polarization vector is always chosen along the rotational
symmetry axis \cite {Com3}.

Figure \ref{Fig1} shows the electronically
selfconsistent (SCF) (fixed geometry) ground state charge densities
for the three different model systems
we consider. Figure \ref{Fig1}(a) shows
the electronic density of spherically symmetric
${\rm Ba@C}_{60}$, Fig.\ref{Fig1}(b) shows the case of
${\rm Ba@C}_{60}({\rm C}_{{\infty}v})$ with Ba displaced from the
center of a spherically symmetric ${\rm C}_{60}$ cage, and
Fig.\ref{Fig1}(c) shows
the symmetric case of  ${\rm Ba@C}_{90}({\rm D}_{{\infty}h})$, where the cage
is a short cylinder with hemispherical ${\rm C}_{60}$ endcaps.
The radius of the Ba 5s/5p charge density is about 2.5 a.u., corresponding
to the radius of the well isolated central charge density
of the ${\rm Ba}^{2+}$ ion core (black spot) in
Figs.(1a)-(1c). In contrast, the Ba 6s-radius is about 6 a.u., causing
large 6s-amplitude beween Ba the ${\rm C}_n$ shell.
The calculated low charge density in this bond region suggests
that the two Ba 6s valence electrons have been transferred
to the ${\rm C}_n$ shell, leaving a doubly ionized central
${\rm Ba}^{2+}$ ion, (Xe ${4d}^{10}5s^25p^6$ configuration), with
some weak effects of bond formation in the displaced case in
Fig.\ref{Fig1}(b).
In the case of ${\rm Ba@C}_{90}$ in
Fig.\ref{Fig1}(c), the strongly ionic character is particularly evident.

The geometries in Fig.\ref{Fig1} do not necessarily represent
equlibrium positions. In fact, we have calculated the variation of
the
total energy for ${\rm Ba@C}_{60}$ as a function
of
displacement of the Ba atom. Technically, the shell is displaced, and
we study the variation of $E_{\text{tot}}({\rm Ba@C}_{60})
- E_{\text{tot}}({\rm C}_{60})$ to minimize systematic errors from
the
finite $l$-expansion. As a result we find that there is a small lowering
of $E_{\text{tot}}$ when the Ba atom is displaced from the center of
${\rm C}_{60}$. Since we have not allowed any freedom in the
shape and
structure of the ${\rm C}_{60}$ shell, we cannot draw any firm
conclusions.
However, our results suggest that here is a tendency for the Ba
atom to
have an off-center equilibrium position, in agreement with
calculations for ${\rm La@C}_{82}$ \cite{6}, ${\rm Ce@C}_{82}$ \cite{7}
and ${\rm Ca@C}_{60}$
\cite{3}. It might even be appropriate to regard the metal atom as
"adsorbed" on the inside of the fullerene cage.

Figure \ref{Fig2} shows a central result of this Letter,
namely the photoabsorption (total photoionization) cross sections
$\sigma(\omega)$ of the ${\rm Ba@C}_n$ clusters described in Fig. 1.
The electric field vector is along the rotational symmetry
axis in displaced ${\rm Ba@C}_{60}({\rm C}_{{\infty}v})$
and along the
cylinder (long) axis in elongated
${\rm Ba@C}_{90}({\rm D}_{{\infty}h})$.

{}From our previous work \cite{WeWa,9} we know that the
environment does not
modify the overall strength and shape of the Ba $4d-{\epsilon}f$ (4d-"4f")
giant dipole
resonance: the effect is to induce a structure which oscillates
around the
atomic "background". The three curves in Fig.\ref{Fig2}
nevertheless present three easily distinguishable patterns -
or "fingerprints" - which may serve as a guide to the structure of
the local environment:

(i) The spherical ${\rm Ba@C}_{60}$ case
(Fig.\ref{Fig1}(a))
is simple: the pronounced XANES oscillations present a clear
signature of the unique distance of a single coordination shell.
These oscillations must be associated with additional nodes of the
continuum ${\epsilon}f$-wave function moving through the
${\rm C}_{60}$ shell into a radial standing wave in the inner
region outside the Ba core. This f-wave function must be of a
molecular,
multiple scattering, kind connected with large-amplitude shape
resonances,
and cannot be described in terms of single scattering.

(ii) The non-spherical, displaced,
${\rm Ba@C}_{60}({\rm C}_{{\infty}v})$ case
(Fig.\ref{Fig1}(b)) lacks pronounced oscillations and,
at first sight, even
characteristic structure seems to be absent.
However, the shape of the 4d cross section is quite different
from that of a free
Ba atom superimposed on a ${\rm C}_{60}$ background \cite{WeWa}
(which corresponds to averaging out the oscillations in
${\rm Ba@C}_{60}$ in case (i) above). With a displaced Ba atom
there are now obviously
two characteristic distances along the electric field vector:
a short "bond" distance of about 5 a.u. and a long distance of
about 8.4 a.u.
A closer inspection of Fig.\ref{Fig2} suggests that
the XANES peaks in case (i) have
been shifted towards higher energies, perhaps due to the a dominating
influence of the shortest ${\rm Ba-C}_{60}$ "bond" distance.

(iii) The non-spherical, but inversion symmetric,
${\rm Ba@C}_{90}({\rm D}_{{\infty}h})$
case (Fig.\ref{Fig1}(c)) lacks pronounced XANES oscillations but
nevertheless shows considerable characteristic structure.
There is no single unique coordination distance, but - as will be
demonstrated below - one can observe structure due to standing waves
along the cylinder (long) axis.

To illuminate the electric-field polarization dependence
of the XANES we compare the cylindrical, cigar-shaped
${\rm Ba@C}_{90}({\rm D}_{{\infty}h})$ with the corresponding
inscribed (short axis) and circumscribed (long axis) spherical systems.
The inscribed system we have already discussed - ${\rm Ba@C}_{60}$.
The circumscribed model system is provided by
a spherically symmetric ${\rm Ba@C}_{130}$ cluster (522 electrons
in the isolated jellium shell).
${\rm Ba@C}_{130}$ is designed to have a unique
coordination length equal to the long axis of ${\rm Ba@C}_{90}$
- the question is whether this similarity will show up in the XANES.

In Fig. \ref{Fig3}, the similarity between
${\rm Ba@C}_{130}$ and ${\rm Ba@C}_{90}$ is striking:
the peak
portion of the ${\rm Ba@C}_{90}$ cross section looks very
similar
to that of ${\rm Ba@C}_{130}$ with
the oscillations averaged out. This suggests that the flat-topped
giant
4d-dipole resonance of ${\rm Ba@C}_{90}$ does reflect the
narrow distribution of coordination distances
defined by the long axis.
Moreover, since ${\rm Ba@C}_{60}$ corresponds to the short
axis and
${\rm Ba@C}_{130}$ to the long axis of ${\rm Ba@C}_{90}$,
Fig.\ref{Fig3} suggests that there will be constructive interference
at the Ba nucleus (4d-shell) around 110 and 140 eV photon energy
and  destructive interference around 120 -130 eV, as indeed seems to be
the case in the ${\rm Ba@C}_{90}$ cross section.

Figure \ref{Fig3} also allows comparison of the XANES
of the spherical ${\rm Ba@C}_{60}$ and ${\rm Ba@C}_{130}$
systems. The immediate impression of the oscillating
structures is that they arise from standing waves inside the
shells, and
that the wavelengths must corespond to the radii of the shells.
Plotted on a final state momentum scale
$k= \sqrt{\epsilon} =\sqrt{\omega+E_{4d}}$,
in both cases the peaks appear as roughly equally spaced
above $\omega=110$ eV.
In the simplest of models we may try to associate the peaks with
high-lying resonances in a radial quantum well, $kR = n\pi$ + const,
$R = \pi/{\delta}k$.
 In this way R(${\rm Ba@C}_{60}$) $\approx$ 7-9 a.u.
and R(${\rm Ba@C}_{130}$) $\approx$ 9-11 a.u., in reasonable
agreement with the given radii of the shells (6.7 and 10 a.u, resp),
but with considerable variations from peak to peak.
Therefore, even if the XANES oscillations look nice and regular, they
cannot be used for precise structure determination in the spirit of
high-energy EXAFS (extended X-ray absorption fine structure).

We are not aware of any experimental results for endohedral
metallofullerene systems that can be compared with the present
theoretical photoabsorption cross sections. However, it is a
fact that
the 4d-giant dipole resonance in the high-$T_c$ superconductor
YBaCuO (Ba in a cage of 6-fold coordinated O; Ba-O distance about
5.1 a.u.) looks very similar to the case of spherically symmetric
${\rm Ba@C}_{60}$ \cite{WeWa} shown in Fig.\ref{Fig1}.
To be able to observe such
strong oscillations in metallofullerens, according to the present
investigation the metal atom must sit at
the center of a spherical cage (unique coordination distance).
Since
this is not likely to occur for ${\rm Ba@C}_{60}$, because
the cage is too large and the Ba atom will sit at an off-center
position,
it might be necessary to
laser evaporate the ${\rm C}_n$ fullerene cage and shrink it
down to, say,
${\rm Ba@C}_{44}$ (in analogy with ${\rm La@C}_{44}$ \cite{4}).
In this case the
cage radius (R $\approx$ 5.7 a.u.) will correspond more closely
to a natural chemical bonding distance -
possibly this is the case that will compare most directly with the
XANES-modulated 4d-giant dipole resonance in YBaCuO.

Finally, in Fig.\ref{Fig4}
we present partial photoionization cross sections
of displaced  ${\rm Ba@C}_{60}({\rm C}_{{\infty}v})$. As expected,
the 4d-f
giant dipole resonance shows upp in all of the $4d,5s,5p$ and
valence level photoemission cross sections with their usual
characteristic
resonance profiles: $n=4$ subshells roughly follow the 4d-f resonance,
while $n{\geq 5}$ subhells (which are basically {\em outside}
the 4d-f dipole)
show their resonance enhancement on the rising edge of the 4d-f giant
dipole resonance.
The weak hump at around 125 eV, discussed above, can be seen
in all of the emission channels.
The valence emission shows resonance enhancement around
100-110 eV at
the same position as the emission from the 5s and 5p orbitals,
and then goes over into "background" emission from the
${\rm C}_{60}$
shell. This indicates significant hybridization of the
${\rm C}_{60}$
shell with Ba 5d orbitals, in line with
the recent photoemisson results for solid
${\rm Ba}_6{\rm C}_{60}$ by Knupfer et al.\cite{Knu,Com3}
and with calculations for ${\rm Ba}_6{\rm C}_{60}$
\cite{ErPe,SaOsh}.

In fact, our complete results show that the
two highest ocupied molecular orbitals (HOMO) have 4f-character
on Ba while the next lower orbitals have 5d- and 6s/6p-character.
This follows in two ways: (i) direct inspection (or partial-wave
analysis) shows that the valence orbital character on Ba is mainly
that of (distorted)
Ba 4f, 5d and 6s/6p; (ii) the partial cross sections show
resonance enhancement of the kind demonstrated in Fig.\ref{Fig4} -
the one for the 4f-hybridized orbitals looks similar to the 4d-cross
section, while the ones for
the 5d- and 6s/6p-hybridized orbitals look similar to the 5s/5p cross
sections. In this way we have a very useful connection between the
character of the initial orbital of the photoemitted electron and
the frequency dependence of the photoelectron intensity.

In conclusion, we have applied a TDLDA  one-center expansion to a
model of a non-spherical metallofullerene cage system
- ${\rm Ba@C}_n$ - in order to study the combined effects of
dynamic screening, multiple scattering and non-spherical local
environment on the  collective Ba 4d-giant dipole resonance
centered around 110 eV photon energy.
For the geometries selected in this work, the XANES
show distinctly different patterns, distinguishing almost
by inspection between center versus off-center position of
the Ba emitter atom as well as spherical versus non-spherical shape
of the cage.
Our results directly concern
photoabsorption and photemission from oriented samples - e.g. solids
or adsorbates -
with polarization vectors along the high-symmetry rotation axis.
However, our analysis suggests that significant parts of
the corresponding XANES will be recognizable also in gas phase spectra.
Whether these XANES signatures are robust enough to
serve as unambiguous "fingerprints" in general situations can only
be decided by comparison with future experimental results.
It would be useful to have photoionization cross sections for
${\rm Xe@C}_n$ \cite{Saunders} and surrounding elements,
e.g. ${\rm Sn@C}_n$-${\rm Ce@C}_n$,
representing different cases of bonding and charge transfer,
shapes of 4d-giant dipole resonances, and many-electron effects.

We would like to thank Zachary H. Levine
and Bo W\"astberg for invaluable support.
We are grateful to R.N. Compton, J. Pendry and Z.C. Ying
for illuminating communications.
This work has been supported by the Swedish Natural Science
Research Council and the National Supercomputer Centre in Sweden.

\begin{figure}
\caption{Self-consistent charge density plots for ${\rm Ba@C}_n$
metallofullerene model clusters (2D jellium model for the $C_n$
nuclear charge; number of electrons chosen to give
closed-shell electronic structure):
(a) ${\rm Ba@C}_{60}$: spherically symmetric ${\rm C}_{60}$ cage
(R=6.7 a.u.; 56+230 electrons) with Ba at the center;
(b) ${\rm Ba@C}_{60}({\rm C}_{{\infty}v})$:
spherically symmetric ${\rm C}_{60}$ cage
(R=6.7 a.u.; 56+230 electrons) with Ba displaced 1.725 a.u. from the center);
(c) ${\rm Ba@C}_{90}({\rm D}_{{\infty}h})$: Ba at the inversion center of
$C_n$ cylinder with
hemispherical $C_{60}$ end caps
($R_{\text{cyl}}=6.7$ a.u., total length=10.0 a.u.;
56+360 electrons).}
\label{Fig1}
\end{figure}

\begin{figure}
\caption{
Photoabsorption cross sections of the ${\rm Ba@C}_n$ clusters defined in Fig.
1:
Spherically symmetric ${\rm Ba@C}_{60}$
$(E_{4d}=$$90.4 eV)$;
displaced Ba, ${\rm Ba@C}_{60}$$({\rm C}_{{\infty}v})$ $(E_{4d}=$$89.6 eV)$;
elongated ${\rm C}_n$, ${\rm Ba@C}_{90}$$({\rm D}_{{\infty}h})$
($E_{4d}=$$91.6 eV$).
For comparison, $E_{4d}=93.7 eV$ in a free LDA Ba atom, and $E_{4d}=$$93.3 eV$
in selfconsistent (SCF) O-Ba-O (the result in Ref.\protect{\cite{9}}
is not SCF). All calculated energies in this work are non-relativistic.
}
\label{Fig2}
\end{figure}

\begin{figure}
\caption{Photoabsorption cross sections of  ${\rm Ba@C}_n$
clusters with an inversion center: Spherical ${\rm Ba@C}_{60}$,
non-spherical ${\rm Ba@C}_{90}({\rm D}_{{\infty}h})$,
and spherical ${\rm Ba@C}_{130}$ ($E_{4d}=94.3 eV$).}
\label{Fig3}
\end{figure}

\begin{figure}
\caption{
Partial photoionization cross sections of
displaced ${\rm Ba@C}_{60}({\rm C}_{{\infty}v})$
(see also Fig.1(b) and Fig.2). The 4d cross section is
the sum of $4d\sigma$, $4d\pi$, and $4d\delta$ components
with energy splitting $\approx 0.016$ eV
($\approx 0.055$ eV in elongated ${\rm Ba@C}_{90}$).
The 5p cross section ($E_{5p}=16.4 eV$) is
the sum of $5p\sigma$ and $5p\pi$ components
with energy splitting $\approx 0.116$ eV.
The $5s\sigma$ level has binding energy $E_{5s}=28.2 eV$.
The valence
cross section represents the integrated emission from
the occupied levels in the $3.14$(HOMO)-$8.60$ eV range below
the vacuum level.
}
\label{Fig4}
\end{figure}


\begin{references}
\bibitem{Com1}	The $L$-expansions for the wave and Green functions
were well converged for $l_{\text{max}} = 14$ in
displaced ${\rm Ba@C}_{60}({\rm C}_{{\infty}v})$
and for $l_{\text{max}} = 17$ in
elongated ${\rm Ba@C}_{90}$$({\rm D}_{{\infty}h})$
(there is no
singularity in the potential of the $C_n$ shell).
\bibitem{Com2}	Corrections to the
dipole approximation will be of order $(R/\lambda)^2$.
In the worst case, the $C_n$ shell, the ratio of the diameter
($R\sim 1$ nm) to the wavelength ($\lambda\sim 10$ nm)
is  $\sim 0.1$.
Therefore, the dipole approximation is an excellent approximation
for the systems considered.
\bibitem{Com3}	Although straightforward, the case of
x/y-polarization (low symmetry direction)
is computationally much more demanding
and beyond the scope of the present work.
\end{references}
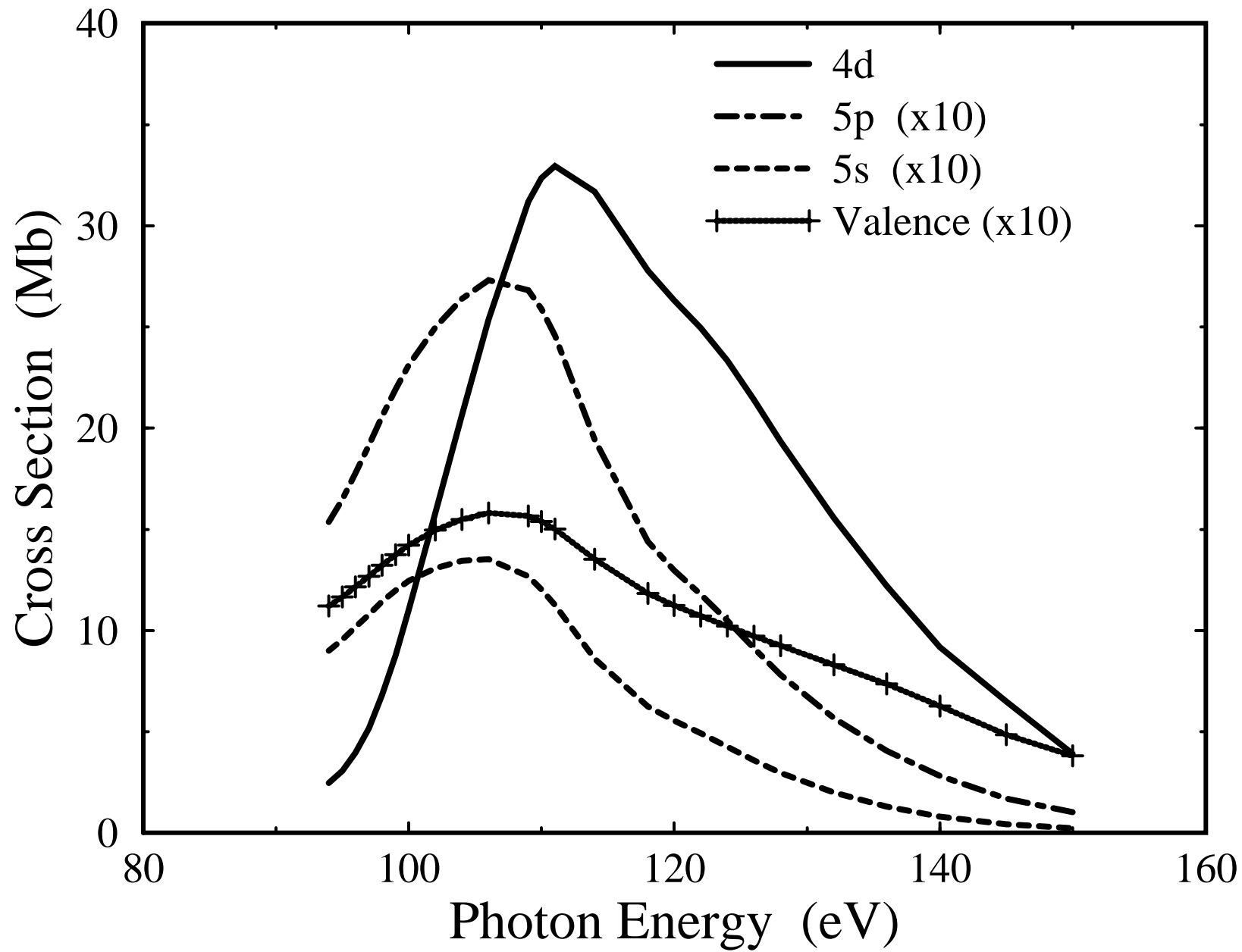
\end{document}